\documentclass[a4paper,11pt]{amsart}
\usepackage{graphicx}
\usepackage{amssymb}
\begin{document}

\hyphenation{gra-vi-ta-tio-nal re-la-ti-vi-ty Gaus-sian
re-fe-ren-ce re-la-ti-ve gra-vi-ta-tion Schwarz-schild
ac-cor-dingly gra-vi-ta-tio-nal-ly re-la-ti-vi-stic pro-du-cing
de-ri-va-ti-ve ge-ne-ral ex-pli-citly des-cri-bed ma-the-ma-ti-cal
de-si-gnan-do-si coe-ren-za pro-blem gra-vi-ta-ting geo-de-sic
per-ga-mon cos-mo-lo-gi-cal gra-vity cor-res-pon-ding
de-fi-ni-tion phy-si-ka-li-schen ma-the-ma-ti-sches ge-ra-de
Sze-keres con-si-de-red tra-vel-ling ma-ni-fold re-fe-ren-ces
geo-me-tri-cal in-su-pe-rable sup-po-sedly at-tri-bu-table
Bild-raum in-fi-ni-tely counter-ba-lan-ces iso-tro-pi-cally
ap-proxi-mate}

\title[On Majorana's equation]
{{\bf On Majorana's equation}}

\author[Angelo Loinger]{Angelo Loinger}
\address{A.L. -- Dipartimento di Fisica, Universit\`a di Milano, Via
Celoria, 16 - 20133 Milano (Italy)}
\author[Tiziana Marsico]{Tiziana Marsico}
\address{T.M. -- Liceo Classico ``G. Berchet'', Via della Commenda, 26 - 20122 Milano (Italy)}
\email{angelo.loinger@mi.infn.it} \email{martiz64@libero.it}

\vskip0.50cm

\begin{abstract}
The physical results of quantum field theory are
\emph{independent} of the various specializations of Dirac's
gamma-matrices, that are employed in given problems. Accordingly,
the physical meaning of Majorana's equation is very dubious,
considering that it is a consequence of \emph{ad hoc} matrix
representations of the gamma-operators. Therefore, it seems to us
that this equation cannot give the equation of motion of the
neutral WIMPs (weakly interacting massive particles), the
hypothesized constitutive elements of the Dark Matter.
\end{abstract}

\maketitle

\vskip0.80cm \noindent \small Key words: SUSY-particles; Dark
Matter. \\ PACS 11.10.Qr -- Relativistic wave-equations.
\normalsize

\vskip1.20cm \noindent \textbf{1.} -- Let us consider the Dirac
equation for the field operator $\psi$ in the absence of any
interaction with other fields:

\begin{equation} \label{eq:one}
i \, \gamma^{\mu} \frac{\partial \psi}{\partial x^{\mu}} - m\psi
=0 \quad,  \quad (\hbar=c=1; \quad \mu=0,1,2,3) \quad ;
\end{equation}

\begin{equation} \label{eq:oneprime}
\gamma^{\mu} \, \gamma^{\nu} + \gamma^{\nu} \, \gamma^{\mu} = 2 \,
\eta^{\, \mu \nu}\tag{1$'$}\quad,
\end{equation}

where $\eta^{\, \mu \nu}$ is the customary Minkowskian tensor.

\par The corresponding charge-conjugate operator $\psi_C$
satisfies the \emph{same} Dirac equation (\ref{eq:one}) \cite{1}.

\par (We recall that with any matrix representation of the
$\gamma^{\mu}$'s, the operator $\psi_C$ is a simple function of
the adjoint of $\psi$).

\par It is clear that Dirac equation (\ref{eq:one}) describes
perfectly also the \emph{neutral} particles with their
\emph{identical} antiparticles.

\vskip1.20cm \noindent \textbf{2.} -- In quantum field theory the
physical concepts (\emph{e.g.}, the energy eigenvalues, the
four-current $j^{\mu}$, \emph{etc.}) and the physical conclusions
are \emph{independent} of any particular matrix representation of
Dirac's $\gamma^{\mu}$--operators. As it is known.

\vskip1.20cm \noindent \textbf{3.} -- Majorana's equation (see
\cite{2} and \cite{3}) is a specialization of Dirac equation
\cite{1} such that the elements of the $\iota^{\mu}$--matrices are
all \emph{imaginary}. Consequently, the expression $[i\,
\gamma^{\mu}_{im} \, \partial_{\mu}-m]$is a \emph{real}
expression, and we can put, with Majorana, $\psi=\varphi + i\chi$,
with a ``real'' $\varphi$ and a `real'' $\chi$ -- two
\emph{selfadjoint} operators --, that satisfy the \emph{same}
Dirac equation. And it seems that Majorana's equation

\begin{equation} \label{eq:two}
i \, \gamma^{\mu}_{im} \frac{\partial \varphi}{\partial x^{\mu}} -
m\varphi =0
\end{equation}

describes particles which coincide with their antiparticles,
\emph{i.e.} neutral objects.

\par However, this conclusion depends on a particular choice of
the $\gamma^{\mu}$--matrices, and is consequently very
problematic. Moreover eq. (\ref{eq:two}) is \emph{not} invariant
under the phase (\emph{gauge}) transformations of $\varphi$:

\begin{displaymath} \label{eq:three}
\left\{ \begin{array}{l} \varphi(x) \rightarrow \varphi'(x)\equiv
\varphi(x) \exp[\pm iF(x)] \quad, \\ \\
\frac{\partial}{\partial x^{\mu}} \rightarrow
\frac{\partial}{\partial x^{\mu}}\mp \frac{\partial F(x)}{\partial
x^{\mu}} \quad. \tag{3}
\end{array} \right.
\end{displaymath}

Of course, if we write

\setcounter{equation}{3}
\begin{equation} \label{eq:four}
i \, \gamma^{\mu}_{im} \frac{\partial \psi}{\partial x^{\mu}} -
m\psi =0 \quad, \quad (\psi=\varphi+i\chi) \quad,
\end{equation}

we have  a standard instance of Dirac equation (\ref{eq:one}),
which is invariant under the phase transformations of $\psi$.

\vskip1.20cm \noindent \textbf{4.} -- According to many
astrophysicists, Dark Matter is composed of WIMPs (weakly
interacting massive particles) \cite{4}. And there is a widespread
belief that these neutral particles are described by Majorana's
equation (\ref{eq:two}. Now, the above considerations show that
this conviction is not well founded.

\par The motions of the hypothetical WIMPs can be properly described by means of Dirac
equation (\ref{eq:one}).

\vskip2.00cm
\begin{center}
\noindent \small \emph{\textbf{APPENDIX A}}
\end{center} \normalsize

\vskip0.40cm \noindent The condition
$\gamma^{\mu}=\gamma^{\mu}_{im}$, $(\mu=0,1,2,3)$, is necessary
and sufficient for the formal validity of Majorana's equation
(\ref{eq:two}). Indeed, let us assume that the $\gamma^{\mu}$'s
are different from the $\gamma^{\mu}_{im}$'s, but -- for
simplicity -- with the same transposition on properties of the
$\gamma^{\mu}_{im}$'s \emph{i.e.}: $\gamma^{0 T}= - \gamma^{0}$;
$\gamma^{kT}=\gamma^{k}$, $(k=1,2,3)$ \cite{3}. We know that,
quite generally, $\gamma^{0}$ is Hermitian and the $\gamma^{k}$'s
are anti-Hermitian.

\par Denoting the adjointness with an asterisk, from the equation

\begin{equation} \label{eq:A1}
\left( i \, \gamma^{0} \partial\,_{0} + i \, \gamma^{k}
\partial\,_{k} - m \right) \psi = 0
\end{equation}

we get easily:

\begin{equation} \label{eq:A2}
\left( i \, \gamma^{0} \partial\,_{0} + i \, \gamma^{k}
\partial\,_{k} - m \right) \psi^{*T} = 0 \quad.
\end{equation}

If $\psi_C$ is the charge-conjugate operator of $\psi$, we have:

\begin{equation} \label{eq:A3}
\left( i \, \gamma^{0} \partial\,_{0} + i \, \gamma^{k}
\partial\,_{k} - m \right) \psi_{C} = 0 \quad.
\end{equation}

Now, the assumption

\begin{equation} \label{eq:A4}
\psi_{C} = \psi^{*T}
\end{equation}

shows that eq. (\ref{eq:A3}) coincides with eq. (\ref{eq:A2}); and
we see that Majorana's condition that $\psi$ is selfadjoint:

\begin{equation} \label{eq:A5}
\psi_{C} = \psi^{*T} = \psi
\end{equation}

implies $\gamma^{\mu} = \gamma^{\mu}_{im}$, \emph{i.e.} the
assumption that the expression $(\ldots)$ of eq. (\ref{eq:A1}) is
\emph{real}.

\par Remak that the restriction $\gamma^{\mu} = \gamma^{\mu}_{im}$
\emph{without} the assumption (\ref{eq:A4}) tells us that $\psi =
\varphi + i \chi$, with $\varphi$ and $\chi$ selfadjoint
operators.

\par It is now obvious to conclude for the validity of Majorana's
equation (\ref{eq:two}). It is not difficult to generalize the
above reasoning for any transposition property of $\gamma^{0},
\gamma^{1}, \gamma^{2}, \gamma^{3}$.

\vskip2.00cm
\begin{center}
\noindent \small \emph{\textbf{APPENDIX B}}
\end{center} \normalsize

\vskip0.40cm \noindent We have considered Majorana's equation,
with Majorana \cite{2} and Wilczek \cite{3}, from the standpoint
of quantum field theory (``second'' quantization of Dirac theory).
Since \emph{ubi maius minus cessat}, the previous treatment
implies that a consideration of Majorana's equation as a
mathematical object of a ``first'' quantization, or of a classical
field theory (wave-picture, \emph{Wellenbild}, in Heisenberg's
terminology) is quite superfluous. However, we think that is is
useful to emphasize the following points.
\begin{enumerate}
    \item[\emph{i})] \emph{``First'' quantization}: the wave-function $\psi$ of
    Schr\"odinger and Dirac equations must be \emph{complex} (see
    Pauli \cite{5}) -- and this fact is sufficient to discard
    Majorana's equation.
    \item[\emph{ii})] \emph{Classical field theory}: the probability density
    and the probability current-density of the ``first''
    quantization assumes a realistic meaning of matter density and
    matter current-density, which involves obviously a
    \emph{non-real} classical field $\psi$: a Majorana's
    equation does not make sense.
\end{enumerate}

\vskip2.00cm
\begin{center}
\noindent \small \emph{\textbf{APPENDIX C}}
\end{center} \normalsize

\vskip0.40cm \noindent We have emphasized in sect. \textbf{1} that
a Dirac neutral particle coincides with its antiparticle. For a
Weyl neutral particle things go otherwise \cite{6}. Indeed, the
charge conjugate of Weyl equation describes an antiparticle, which
is characterized by the opposite sign of particle \emph{helicity}
$\vec{\sigma} \cdot \vec{p} \, / |\vec{p}|$. This unique
difference between particle and antiparticle has only a
mathematical origin: Weyl equation is not invariant with respect
to space reflections.

\end{document}